\documentclass{llncs}
\usepackage{graphicx}
\usepackage{subfigure}
\usepackage{algorithm}
\usepackage{algorithmic}

\newcommand{\pardef}[1]{\textbf{#1}\quad}
\newcommand{\BREAK}{\textbf{break}}
\renewcommand{\algorithmicrequire}{\textbf{Input:}}
\renewcommand{\algorithmicensure}{\textbf{Output:}}
\renewcommand{\algorithmiccomment}[1]{/* #1 */}

\begin{document}
\title{Novel algorithm to calculate hypervolume indicator of Pareto approximation set}
\author{Qing Yang\inst{1} \and Shengchao Ding\inst{2,3}
\institute{School of Computer Science and
Technology, South-Central University for Nationalities, Wuhan, China \and Institute of
Computing Technology, Chinese Academy of Sciences, Beijing, China \and Graduate
University of the Chinese Academy of Sciences,
Beijing, China\\
 \email{dingshengchao@ict.ac.cn}}}%
\maketitle

\begin{abstract}
Hypervolume indicator is a commonly accepted quality measure for comparing Pareto
approximation set generated by multi-objective optimizers. The best known algorithm to
calculate it for $n$ points in $d$-dimensional space has a run time of $O(n^{d/2})$ with
special data structures. This paper presents a recursive, vertex-splitting algorithm for
calculating the hypervolume indicator of a set of $n$ non-comparable points in $d>2$
dimensions. It splits out multiple child hyper-cuboids which can not be dominated by a
splitting reference point. In special, the splitting reference point is carefully chosen
to minimize the number of points in the child hyper-cuboids. The complexity analysis
shows that the proposed algorithm achieves $O((\frac{d}{2})^n)$ time and $O(dn^2)$ space
complexity in the worst case.
\end{abstract}

\section{Introduction}
Optimization for multiple conflicting objectives results in more than one optimal
solutions (known as Pareto-optimal solutions). Although one of these solutions is to be
chosen at the end, the recent trend in evolutionary and classical multi-objective
optimization studies have focused on approximating the set of Pareto-optimal solutions.
However, to assess the quality of Pareto approximation set, special measures are
needed~\cite{Zitzler03}.

Hypervolume indicator is a commonly accepted quality measure for comparing approximation
set generated by multi-objective optimizers. The indicator measures the hypervolume of
the dominated portion of the objective space by Pareto approximation set and has
received more and more attention in recent
years~\cite{Zitzler98,Zitzler99c,Zitzler03,Zitzler07Hypervolume}.

There have been some studies that discuss the issue of fast hypervolume
calculation~\cite{WFG2005d,While2006Hypervolume,FPL06Hypervolume,BeumeRudolph06}. These
algorithms partition the covered space into many cuboid-shaped regions, within which the
approach considering the dominated hypervolume as a special case of Klee's measure
problem is regarded as the current best one. This approach~\cite{BeumeRudolph06} adopts
orthogonal partition tree which requires $O(n^{d/2})$ storage and streaming
variant~\cite{Edels85}. Conceptual simplification of the implementation are concerned
and thus the algorithm achieves an upper bound of $O(n\log{n}+n^{d/2})$ for the
hypervolume calculation. Ignoring the running time of sorting the points according to
the $d$-th dimension, $O(n\log{n})$, the running time of this approach is exponential of
the dimension of space $d$.

This paper develops novel heuristics for the calculation of hypervolume indicator.
Special technologies are applied and the novel approach yields upper bound of
$O((\frac{d}{2})^n)$ runtime and consumes $O(dn^2)$ storage. The paper is organized as
follows. In the next section, the hypervolume indicator is defined, and some background
on its calculation is provided. Then, an algorithm is proposed which uses the so-called
vertex-splitting technology to reduce the hypervolume. The complexities of the proposed
algorithm are analyzed in Section~\ref{sec:complexity}. The last section concludes this
paper with an open problem.

\section{Background}
Without loss of generality, for multi-objective optimization problems, if the $d$
objective functions $f=(f_1,\ldots,f_d)$ are considered with $f_i$ to be minimized, not
one optimal solution but a set of good compromise solutions are obtained since that the
objectives are commonly conflicting. The compromise solutions are commonly called Pareto
approximation solutions and the set of them is called the Pareto approximation set. For
a Pareto approximation set $M=\{y_1,y_2,\ldots,y_n\}$ produced in a run of a
multi-objective optimizer, where $y_i=(y_{i1},\ldots,y_{id})\in M \subset \mathbf{R}^d$,
all the solutions are non-comparable following the well-known concept of Pareto
dominance. Specially, we say that $y_i$ dominates $y_k$ at the $j$-th dimension if
$y_{ij}<y_{kj}$.

\begin{figure}
\centering \subfigure[A hypervolume indicator in the two-objective
case]{\label{fig:2d-set}
\includegraphics[width=0.45\textwidth]{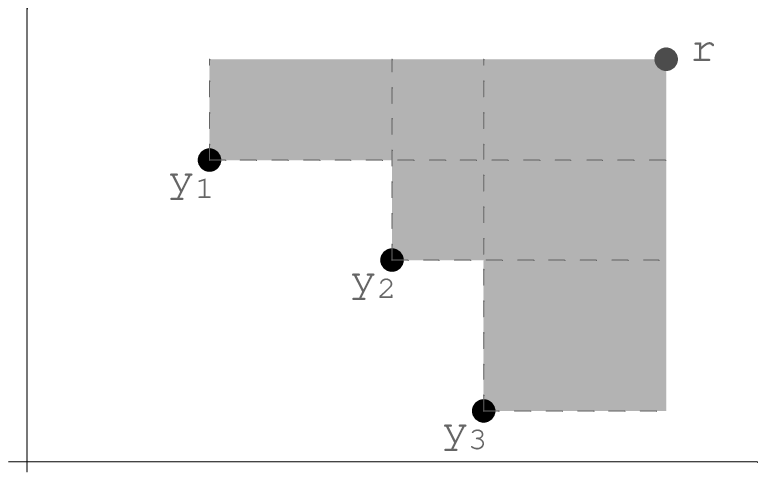}}\qquad
\subfigure[A hypervolume indicator in the three-objective case. To lay out the cuboids
well, the axes are rotated where the reference point is shaded]{\label{fig:3d-set}
\includegraphics[width=0.45\textwidth]{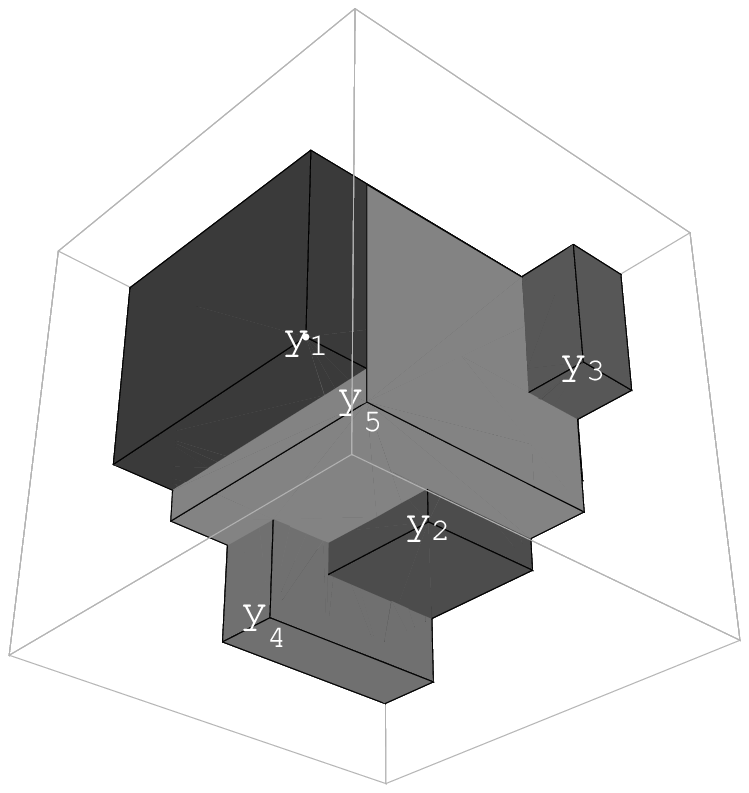}}
\end{figure}

The unary hypervolume indicator of a set $M$ consists of the measure of the region which
is simultaneously dominated by $M$ and bounded above by a reference point
$r=(r_1,\ldots,r_d)\in \mathbf{R}^d$ such that $r_j \geq
\max_{i=1,\ldots,n}{\{y_{ij}\}}$. In the context of hypervolume indicator, we call the
solutions in $M$ as the dominative points. As illustrated in Fig.~\ref{fig:2d-set}, the
shading region consists of an orthogonal polytope, and may be seen as the union of three
axis-aligned hyper-rectangles with one common vertex, i.e., the reference point $r$.
Another example in three dimensional space is shown in Fig.~\ref{fig:3d-set}, where five
dominative points, $y_1=(1,2,3),y_2=(4,3,2),y_3=(5,1,4),y_4=(3,5,1),y_5=(2,2,2.5)$, and
the reference point $r=(6,6,6)$ are considered. The volume is the union of the volumes
of all the cuboids each of which is bounded by a vertex, where the common regions are
counted only once. If a point $y_k$ is dominated by another point $y_i$, the cuboid
bounded by $y_k$ is completely covered by the cuboid bounded by $y_i$. And thus only the
non-dominated points contribute to the hypervolume.

\section{The proposed algorithm}
In other works, e.g. the work of Beume and Rudolph~\cite{BeumeRudolph06}, the
hyper-cuboid in $d$-dimensional space are partitioned into child hyper-cuboids along the
$d$-th dimension and then all these child hypervolumes are gathered together by the
inclusion-exclusion principle~\cite{Overmars91}.

In this paper, we step in another way. The hyper-cuboid is partitioned into child
hyper-cuboids at some splitting reference points and then all the child hypervolumes are
gathered directly. More detailed, given a point $y_i\in M$, each of other points in $M$
must dominated $y_i$ at some dimensions for the non-comparable relation. If the parts
over $y_i$ are handled, the problem of calculating the hypervolume bounded by $M$ and
the reference point is figured out. The additional part partitioned out at the $j$-th
dimension is also a $d$-dimensional hyper-cuboid whose vertices are ones beyond $y_i$ at
such dimension. Their projections on the hyperplane orthogonal to dimension $j$ are all
dominated by $y_i$, and thus are free from consideration. It should be noted that the
reference point of child hyper-cuboid is altered to $r' =
(r_1,\ldots,y_{ij},\ldots,r_d)$, namely the $j$-th coordinate is replaced by $y_{ij}$.
The other child hyper-cuboids are handled in the similar way. In these processes, the
given point is called the splitting reference point.

Obviously, the hyper-cuboids with more dominative points require more run time to
calculate the hypervolumes. To reduce the whole run time for calculating all these child
hyper-cuboids, the splitting reference point should be carefully selected. The strategy
adopted in this paper is described as follows.
\begin{description}
\item (1) Let $k=n-1$ and choose a point with the least dimensions on
which the point dominated by other $k$ points.
\item (2) If some points tie, update $k$ as $k-1$ and then within these points, choose a point
with the least dimensions on which the point dominated by other $k$ points.
\item (3) Repeat the similar process until only single point is left or $k=1$. And if
$k=1$ and several points are left, the first found point is selected.
\end{description}
By the above principle, as an example, not $y_2$ or other points but $y_5$ is chosen as
the first splitting reference point for the case shown in Fig.~\ref{fig:3d-set}. Two
child cuboids each bounded by one points and another child cuboid bounded by two points
are generated by splitting along $y_5$. This is the optimal strategy in such case.

The algorithm to calculate the hypervolume is shown in
Algorithm~\ref{algo:calcHypervolume}. Some major parameters are as follows.
\begin{itemize}
\item \pardef{int[n][d] order} The orders of all the dominative points at each dimension are
represented by a two-dimensional array of integer.
\item \pardef{int split} The index of the point at which the hyper-cuboid is cut to
generate multiple child hyper-cuboids is called $split$.
\item \pardef{int[n] splitCount} The numbers of $k$ present in the $split$-th row of the array
$order$ are saved in $splitCount$, where $k=0,\ldots,n-1$.
\item \pardef{int[n] coveredCount} The numbers of $k$ present in the current
checked row of the array $order$ are save in $coveredCount$, where $k=0,\ldots,n-1$.
\end{itemize}
Moreover, some conventions are explained as follows.
\begin{itemize}
\item The subscript of $y_{ij}$ begins with $1$ while the index of array begins with
$0$. Thus $y_{ij}$ is same as $y[i-1][j-1]$.
\item Assume $a$ and $b$ are two arrays and $n$ is an element. $a[]\Leftarrow n$ means
    setting each element of $a$ as $n$, while $a[ ] \Leftarrow b[ ]$ means copying
    all the elements of $b$ to $a$ pairwise.
\item Assume $S$ is a set and $x$ is an element. $S\Leftarrow S+\{x\}$ means appending
    a copy of $x$ to $S$.
\end{itemize}

The inputs of the algorithm are a set of non-dominated (dominative) points and a
reference point, thus the hyper-cuboids are represented implicitly.

\begin{algorithm}
\caption{Calculate Hypervolume, $CalcVolume(H)$}\label{algo:calcHypervolume}
\begin{algorithmic}[1]
\REQUIRE The hyper-cuboid $H$ defined by the dominative points $\{y_1,y_2,\ldots,y_n\}$
where $y_i = (y_{i1},\ldots,y_{id})$,
    and the reference point $r=(r_1,\ldots,r_d)$, namely $H=\{y_1,\ldots,y_n,r\}$.
    The initial number $n$ of dominative points can be obtained from the length of $H$
    and the dimension $d$ is known too.%
\ENSURE The hypervolume of $H$, $volume$.%
\STATE \COMMENT{initialization}%
\IF{$n=1$}%
    \RETURN $\prod_{j=1}^{d}{|r_j-y_{1j}|}$;
\ENDIF%
\STATE $volume \Leftarrow 0$;%
\STATE $splitCount[] \Leftarrow n$;%
\STATE \COMMENT{count the numbers of points dominating every point at each dimension}
\FOR{$j=1$ to $d$}%
\STATE sort $y_{1j},\ldots,y_{nj}$;%
\FORALL{$i$ such that $1\leq i \leq n$}%
\STATE $order[i-1][j-1] \Leftarrow $ number of points dominating $y_{ij}$ strictly;%
\ENDFOR
\ENDFOR%
\STATE \COMMENT{estimate $split$ based on the statistical results of $order$}%
\FOR{$i=1$ to $n$} %
    \STATE $coveredCount[] \Leftarrow 0$;%
    \FOR{$j=1$ to $d$}%
    \STATE $coveredCount[order[i-1,j-1]]$++;%
    \ENDFOR%
    \FOR{$k=n-1$ downto $0$}%
    \IF{$coveredCount[k]<splitCount[k]$}%
    \STATE $split \Leftarrow i$;%
    \STATE $splitCount[] \Leftarrow coveredCount[]$;
    \STATE \BREAK;
    \ENDIF%
    \ENDFOR%
\ENDFOR %
\STATE \COMMENT{cut $H$ at each dimension through the point indexed by $split$}%
\FOR{$j=1$ to $d$}%
    \IF{$order[split-1][j-1]>0$}%
        \STATE $H2 \Leftarrow \{\}$;%
        \FORALL{$y_i$ in $H\backslash \{y_{split},r\}$}%
            \IF{$y_{ij}$ is dominated strictly by $y_{split,j}$}%
                \STATE $H2 \Leftarrow H2 + \{y_i\}$;%
                \STATE $y_{ij} \Leftarrow y_{split,j}$;%
            \ENDIF%
            \STATE \COMMENT{Here $y_i$ can be removed from $H$ if $y_i$ is dominated strictly by $y_{split}$}%
        \ENDFOR%
        \STATE $r2 \Leftarrow r$;%
        \STATE $r2[j-1] \Leftarrow y_{split,j}$;%
        \STATE $H2 \Leftarrow H2 + \{r2\}$;%
        \STATE $volume \Leftarrow volume + CalcVolume(H2)$;%
    \ENDIF%
\ENDFOR%
\STATE $volume \Leftarrow volume + \prod_{j=1}^{d}{|r_j-y_{split,j}|}$;%
\RETURN $volume$;
\end{algorithmic}
\end{algorithm}

In fact, when the hyper-cuboid is cut into two child hyper-cuboids, there may be some
points dominated by the splitting reference point in the bigger cuboid, and thus such
points could be removed from the points set $H$. In the proposed algorithm, it does not
matter whether those points are removed or not.

\section{Complexity Analysis}\label{sec:complexity}
Before discussing the time-space complexity of the proposed algorithm, some properties
are presented firstly.
\begin{lemma}\label{lemma:delta}
Let $\delta_{ij}$ be the number of points dominating $y_i$ at the $j$-th dimension. Then
\begin{description}
\item (1) For $d\geq 2$ and each $i\in \{1,\ldots,n\}$, $\sum_{j=1}^{d}{\delta_{ij}} \geq n-1$.
\item (2) For $d\geq 2$ and each $j\in \{1,\ldots,d\}$,
    $\sum_{i=1}^{n}{\delta_{ij}} \leq \frac{n}{2}(n-1)$.
\item (3) For $d = 2$ and each $i\in \{1,\ldots,n\}$, $\sum_{j=1}^{d}{\delta_{ij}} = n-1$.
\item (4) $ \sum_{i=1}^{n}{\sum_{j=1}^{d}{\delta_{ij}}} \leq \frac{dn}{2}(n-1)$.
\item (5) For $d\geq 2$ and each $i\in \{1,\ldots,n\}$, $\sum_{j=1}^{d}{\delta_{ij}}\leq \frac{d}{2}(n-1)$.
\end{description}
\end{lemma}
\begin{proof}
It is clear that (2) $\Rightarrow$ (4) $\Rightarrow$ (5). The follows show (1), (2) and
(3).
\begin{description}
\item (1) (By Contradiction.) Assume to the contrary there is some $i\in
\{1,\ldots,n\}$, $\sum_{j=1}^{d}{\delta_{ij}}<n-1$. If this is the case, there are at
least one $y_k$ where $k\neq i$ such that each $y_{ij}$ dominates $y_{kj}$ for all $j\in
\{1,\ldots,d\}$. It follows that $y_i$ dominates $y_k$, which contradicts our assumption
that all the elements in $\{y_1,\ldots,y_n\}$ are non-comparable.

\item (2) Given $j$, sort all $y_{ij}$ where $i=1,\ldots,n$ and label each $y_{ij}$ a sequence
number $I(i)$ which ranges from 0 to $n-1$. Thus $\sum_{i=1}^{n}{I(i)} =
\frac{n}{2}(n-1)$. There are two cases to consider. Firstly, if all $y_{ij}$ are
different each other, then $\delta_{ij} = I(i)$. It follows that
$\sum_{i=1}^{n}{\delta_{ij}} = \frac{n}{2}(n-1)$. Secondly, if there are same elements
within $\{y_{1j},\ldots,y_{nj}\}$, without loss of generality, suppose $y_{ij}= y_{kj}$
and $I(k) = I(i)+1$. Then $\delta_{ij} = \delta_{kj} = I(i)<I(k)$, it follows that
$\sum_{i=1}^{n}{\delta_{ij}}<\frac{n}{2}(n-1)$. This completes the proof.

\item (3) (By contradiction.) For any $y_i$, $\sum_{j=1}^{d}{\delta_{ij}}<n-1$ is excluded by (1) of this lemma.
Thus $\sum_{j=1}^{d}{\delta_{ij}}>n-1$ for some $y_i$ is considered. If this is the
case, we obtain $\sum_{i=1}^{n}{\sum_{j=1}^{d}{\delta_{ij}}} > n(n-1)$, contradicting
(2) of this lemma, which implies $\sum_{i=1}^{n}{\sum_{j=1}^{2}{\delta_{ij}}} =
\sum_{j=1}^{2}{\sum_{i=1}^{n}{\delta_{ij}}}\leq n(n-1)$, namely
$\sum_{i=1}^{n}{\sum_{j=1}^{d}{\delta_{ij}}} \leq n(n-1) $.
\end{description}
\end{proof}

\begin{lemma}\label{lemma:omega}
Let $\omega_i(k)$ be the amount of $k$ in all $\delta_{ij}$ where $j=1,\ldots,d$, namely
$
    \omega_i(k) = |\{j:\delta_{ij}=k,j=1,\ldots,d\}|
$. Then
\begin{description}
\item (1) $0\leq \omega_i(k) \leq d$ for any $i$ and $k$;
\item (2) $\sum_{i=1}^{n}{\omega_i(k)} \leq d$ for any $k$;
\item (3) $\sum_{k=0}^{n-1}{k \omega_i(k)} \leq \frac{d}{2}(n-1)$  for any $i$.
\end{description}
\end{lemma}
\begin{proof}
By the definition of $\omega_{i}(k)$, it is clear that all statements follows
Lemma~\ref{lemma:delta}.
\end{proof}

\begin{lemma}\label{lemma:runtime}
Let $f(n,d)$ be the runtime of Algorithm~\ref{algo:calcHypervolume} to compute a
hypervolume with $n$ dominative points in a $d$-dimensional space. Then
\begin{description}
\item (1) $f(n,d)+f(m,d) > f(n-1,d)+f(m+1,d)$ where $n-m>1$;
\item (2) $f(n,d) > f(m,d) + f(n-m,d)$ where $n>m$;
\item (3) $f(n,d)$ is minimal when $\sum_{j=1}^{d}{\delta_{ij}}  = n-1$ and
$|\delta_{ij}-\delta_{ik}|\leq 1$ for any $j$ and $k$;
\item (4) $f(n,d)$ is maximal when $\sum_{j=1}^{d}{\delta_{ij}} =
\frac{d}{2}(n-1)$ for any $i$ and $\omega_i(k) = \frac{d}{n}$ for any $i$ and each
$k=0,\ldots,n-1$.
\end{description}
\end{lemma}
\begin{proof}
\begin{description}
\item (1) and (2) are clear.
\item (3) By the process of Algorithm~\ref{algo:calcHypervolume}, given some $i$,
\begin{equation}\label{eqn:fnd1}
    f(n,d) = dn\log n + \sum_{j=1}^{d}{f(\delta_{ij},d)}
\end{equation}
By (1) of Lemma~\ref{lemma:delta}, $\sum_{j=1}^{d}{\delta_{ij}} \geq n-1$. It is clear
that for a given $i$, it is necessary that $\sum_{j=1}^{d}{\delta_{ij}}=n-1$ to minimize
$f(n,d)$. In addition, all the $\delta_{ij}$ must share alike, i.e.
$|\delta_{ij}-\delta_{ik}|\leq 1$ for any $j$ and $k$. If this is not the truth, suppose
$\delta_{ij} - \delta_{ik} >1$. Thus by (1) of this lemma,
\begin{equation}
    f(\delta_{ij},d) + f(\delta_{ik},d) > f(\delta_{ij}-1,d) + f(\delta_{ik}+1,d)
\end{equation}
Let $\delta_{ij'} = \delta_{ij} -1$ and $\delta_{ik'} = \delta_{ik}+1$. $\delta_{ij'}$
and $\delta_{ik'}$ can be modified in the similar way until $|\delta_{ij} - \delta_{ik}|
\leq 1$. This completes the proof.

\item (4) By (5) of Lemma~\ref{lemma:delta}, $\sum_{j=1}^{d}{\delta_{ij}} \leq \frac{d}{2}(n-1)$. It is clear
that for a given $i$, it is necessary that $\sum_{j=1}^{d}{\delta_{ij}} =
\frac{d}{2}(n-1)$ to maximize $f(n,d)$.

Hence Eqn.~(\ref{eqn:fnd1}) is written as follows,
\begin{equation}\label{eqn:fnd2}
    f(n,d) = dn\log n + \sum_{k=1}^{n-1}{\omega(k)f(k,d)}
\end{equation}
Suppose $y_i$ is the splitting reference point chosen by
Algorithm~\ref{algo:calcHypervolume}, $\omega_i(n-1) \leq \frac{d}{n}$, or else
contradicting $\sum_{i=1}^{n}{\omega_i(n-1)} \leq d$. To maximize $f(n,d)$ in
Eqn.~(\ref{eqn:fnd2}), let $\omega_i(n-1) = \frac{d}{n}$. Similarly, we get
$\omega_i(n-2) = \frac{d}{n}$, $\ldots$, $\omega_i(1) = \frac{d}{n}$, and so on. It is
exactly $\sum_{k=0}^{n-1}{k \omega_i(k)} = \frac{d}{2}(n-1)$. This completes the proof.

\end{description}
\end{proof}

\subsection{Bounds of runtime at special cases}
First of all, it is clear that $f(1,d) = d$. By (3) of Lemma~\ref{lemma:runtime}, the
algorithm performs best when each $\delta_{ij}$ shares alike for the chosen $i$. If
$d\geq n-1$, $\delta_{ij}\leq 1$ for any $j$. Thus
\begin{equation}
    f(n,d)  = dn\log n + (n-1)f(1,d)
\end{equation}
which implies $f(n,d) = \Omega(dn\log n)$. If $d< n-1$, $\delta_{ij}> 1$ for any $j$. In
the rough, we get
\begin{equation}\label{eqn:lowerRuntime}
 f(n,d) = dn\log{n} + d \cdot
f(\frac{n-1}{d}, d) < dn\log{n} + d \cdot f(\frac{n}{d},d)
\end{equation}
It can be obtained from Eqn.~(\ref{eqn:lowerRuntime}) that $f(n,d) =
\Theta(dn\log{n}\log_{d}{n})$ even when $f(\frac{n-1}{d},d)$ is relaxed to
$f(\frac{n}{d},d)$.

Fredman and Weide~\cite{FredmanWeide78} have shown that Klee's measure problem has a
lower bound of $\Omega(n\log{n})$ for arbitrary $d\geq 1$. Just as Beume and
Rudolph~\cite{BeumeRudolph06} have mentioned, although it is unknown what the lower
bound for calculating the hypervolume is, it is definitely not harder than solving KMP
because it is a special case of KMP. Therefore, there is a gap between the lower bound
of the proposed algorithm and the actual lower bound of calculating the hypervolume.

In the average cases, suppose that for the given splitting reference point $y_i$,
$\sum_{j=1}^{d}{\delta_{ij}} = \frac{d}{2}(n-1)$. Meanwhile, each $\delta_{ij}$ shares
alike, i.e. $\delta_{ij} = \frac{n-1}{2}$. Thus,
\begin{equation}
f(n,d) = dn\log{n} + d\cdot f(\frac{n-1}{2},d)< dn\log{n} + d\cdot f(\frac{n}{2},d)
\end{equation}
which implies the runtime of the proposed algorithm is $\Theta(dn^{\log{d}})$ at the
given cases.

\subsection{Upper bound of runtime}
By (2) of Lemma~\ref{lemma:runtime}, $f(n-1)>f(n-1-k)+f(k)$ for any
$k=1,\ldots,\frac{n-2}{2}$. And by (4) of Lemma~\ref{lemma:runtime}, at the worst cases,
we have
\begin{equation}
\begin{array}{lcl}
f(n,d)& = & dn\log{n} + \frac{d}{n}\left( f(n-1)+ f(n-2) + \ldots + f(2) + f(1)\right)
\\
& < & dn\log{n} + \frac{d}{n}(1+\frac{n-2}{2})f(n-1)\\
& < & dn\log{n} + \frac{d}{2}f(n-1)
\end{array}
\end{equation}
which implies that the proposed algorithm for computing the hypervolume bounded by $n$
points and a reference point in $d$-dimensional space has a runtime of
$O((\frac{d}{2})^n)$ at the worst cases.

\subsection{Space complexity}
Let $g(n,d)$ be the used storage by Algorithm~\ref{algo:calcHypervolume}. In the
proposed algorithm, every child hypervolume is calculated one by one. Since the storage
can be reused after the former computation has been completed, $g(n,d)$ is only related
to the maximum usage of all the computations of child hypervolumes. Hence,
\begin{equation}
g(n,d) = dn + \max_{i\in \{1,\ldots,n\},j\in \{1,\ldots,d\}}{\{g(\delta_{ij},d):0 \leq
\delta_{ij} \leq n-1\}}
\end{equation}
Thus the upper bound of space is as follows.
\begin{equation}
g(n,d) = dn+ g(n-1,d)
\end{equation}
where $g(1)=d$. It is easy to obtain an $O(dn^2)$ space upper bound for the proposed
algorithm.

Combining the above analyses together, we obtain the time-space complexity of the
proposed algorithm.
\begin{theorem}
The hypervolume of a hyper-cuboid bounded by $n$ non-comparable points and a reference
point in $d$-dimensional space can be computed in time $O((\frac{d}{2})^n)$ using
$O(dn^2)$ storage.
\end{theorem}

\section{Conclusions}
A fast algorithm to calculate the hypervolume indicator of Pareto approximation set is
proposed. In the novel algorithm, the hyper-cuboid bounded by non-comparable points and
the reference point is partitioned into many child hyper-cuboids along the carefully
chosen splitting reference point at each dimension. The proposed approach is very
different to the technique used in other works where the whole $d$-dimensional volume is
calculated by computing the $(d-1)$-dimensional volume along the dimension $d$. Such
difference results in very different time bounds, namely $O((\frac{d}{2})^n)$ for our
work and $O(n^{\frac{d}{2}})$ for the best previous result. Neither kind of technique
can exceed the other completely and each has his strong point. Additionally, the amount
of storage used by our algorithm is only $O(dn^2)$ even no special technique is
developed to reduce the space complexity.

As the context has mentioned, it is very important to choose appropriate splitting
reference point for our algorithm. Well selected point can reduce number of points in
separated parts and thus cut down the whole runtime. We do not know whether the strategy
adopted in this paper is optimal or near optimal. Further investigations should be
worked on.


\begin{thebibliography}{10}

\bibitem{Zitzler03}
Zitzler, E., Thiele, L., Laumanns, M., Fonseca, C.M., da~Fonseca, V.G.:
\newblock Performance {A}ssessment of {M}ultiobjective {O}ptimizers: {A}n
  {A}nalysis and {R}eview.
\newblock IEEE Transactions on Evolutionary Computation \textbf{7}(2) (2003)
  117--132

\bibitem{Zitzler98}
Zitzler, E., Thiele, L.:
\newblock Multiobjective {O}ptimization {U}sing {E}volutionary {A}lgorithms ---
  {A} {C}omparative {S}tudy.
\newblock In Eiben, A.E., ed.: Parallel Problem Solving from Nature V,
  Amsterdam, Springer-Verlag (1998)  292--301

\bibitem{Zitzler99c}
Zitzler, E., Thiele, L.:
\newblock Multiobjective {E}volutionary {A}lgorithms: {A} {C}omparative {C}ase
  {S}tudy and the {S}trength {P}areto {A}pproach.
\newblock IEEE Transactions on Evolutionary Computation \textbf{3}(4) (1999)
  257--271

\bibitem{Zitzler07Hypervolume}
Zitzler, E., Brockhoff, D., Thiele, L.:
\newblock The hypervolume indicator revisited: On the design of
  pareto-compliant indicators via weighted integration.
\newblock In: Proceedings of the 4th International Conference on Evolutionary
  Multi-Criterion Optimization (EMO 2007). Volume 4403., Springer-Verlag (2007)
   862--876

\bibitem{WFG2005d}
While, L., Bradstreet, L., Barone, L., Hingston, P.:
\newblock Heuristics for optimising the calculation of hypervolume for
  multi-objective optimisation problems.
\newblock In: The 2005 IEEE Congress on Evolutionary Computation. Volume~3.
  (2005)  2225--2232

\bibitem{While2006Hypervolume}
While, L., Hingston, P., Barone, L., Huband, S.:
\newblock A faster algorithm for calculating hypervolume.
\newblock IEEE Transactions on Evolutionary Computation \textbf{10}(1) (2006)
  29--38

\bibitem{FPL06Hypervolume}
Fonseca, C.M., Paquete, L., {n}ez, M.L.I.:
\newblock An improved dimension-sweep algorithm for the hypervolume indicator.
\newblock In: {IEEE} Congress on Evolutionary Computation (CEC 2006). (2006)
  1157--1163

\bibitem{BeumeRudolph06}
Beume, N., Rudolph, G.:
\newblock Faster s-metric calculation by considering dominated hypervolume as
  klee's measure problem.
\newblock In Kovalerchuk, B., ed.: Proceedings of the Second IASTED Conference
  on Computational Intelligence, Anaheim, ACTA Press (2006)  231--236

\bibitem{Edels85}
Edelsbrunner, H., Overmars, M.H.:
\newblock Batched dynamic solutions to decomposable searching problems.
\newblock Journal of Algorithms \textbf{6}(4) (1985)  515--542

\bibitem{Overmars91}
Overmars, M.H., Yap, C.K.:
\newblock New upper bounds in klee's measure problem.
\newblock SIAM Journal on Computing \textbf{20}(6) (1991)  1034--1045

\bibitem{FredmanWeide78}
Fredman, M.L., Weide, B.:
\newblock The complexity of computing the measure of $\bigcup{[a_i,b_i]}$.
\newblock Communications of ACM \textbf{21} (1978)  540--544

\end{thebibliography}

\end{document}